\documentclass[12pt]{article}
\usepackage{sigsam-edit, amsmath}
\usepackage{enumerate}
\usepackage{shortlst}
\usepackage{color}

\usepackage{graphicx}
\usepackage{booktabs}

\issue{}
\articlehead{}

\titlehead{A comparison of three heuristics to choose the variable ordering for CAD}
\authorhead{Zongyan Huang, Matthew England, David Wilson, James H. Davenport and Lawrence C. Paulson}
\begin{document}

\title{A comparison of three heuristics to choose the variable ordering for cylindrical algebraic decomposition}

\author{Zongyan Huang$^1$, Matthew England$^2$, David Wilson$^2$,\\ 
James H. Davenport$^2$ and Lawrence C. Paulson$^1$ \\
$^1$ University of Cambridge Computer Laboratory, Cambridge CB3 0FD, U.K. \\
\url{{zh242,lp15}@cam.ac.uk} \\
$^2$ Department of Computer Science, University of Bath, Bath, BA2 7AY, U.K. \\
\url{{M.England,D.J.Wilson,J.H.Davenport}@bath.ac.uk}
}

\date{}

\maketitle

\begin{abstract}
Cylindrical algebraic decomposition (CAD) is a key tool for problems in real algebraic geometry and beyond.  When using CAD there is often a choice over the variable ordering to use, with some problems infeasible in one ordering but simple in another.  Here we discuss a recent experiment comparing three heuristics for making this choice on thousands of examples.
\end{abstract}

\section{Background}
A cylindrical algebraic decomposition (CAD) dissects $\mathbb{R}^n$ into cells, each described by polynomial relations and arranged cylindrically so that the projection of any two cells into lower coordinates is equal or disjoint.  CAD is a key tool, both for its original motivation of quantifier elimination (QE) over real-closed fields and many other applications discovered since.  For a more detailed introduction see Huang \textit{et al}. \cite{HEWDPB14} and the references within.  When using CAD there may be a choice for the variable ordering and it is well known that this can have a great effect on the tractability of a problem.  
We recently tested the following three heuristics for picking this ordering.
\begin{description}
\item[Brown:] 
Suggested by Brown \cite{Brown2004}, this heuristic chooses a variable ordering according to three criteria on the input system.  We start with the first and break ties with successive ones.  The advice is to eliminate a variable first if:
\vspace*{-10pt}
\begin{shortenumerate}[(1)]
\item it has lower overall degree in the input;
\item it has lower (maximum) total degree of those terms in the input in which it occurs;
\item there is a smaller number of terms in the input which contain the variable.
\end{shortenumerate}
\vspace*{-10pt}
\item[sotd:] 
Suggested by Dolzmann \textit{et al}. \cite{DSS04}, this heuristic constructs the full set of projection polynomials for each ordering and selects the ordering whose set has the lowest \emph{sum of total degree} for each of the monomials in each of the polynomials. 
\vspace*{-5pt}
\item[ndrr:] 
Suggested by Bradford \textit{et al}. \cite{BDEW13}, this heuristic also constructs the full projection set, and selects the one with the lowest \emph{number of distinct real roots} of the univariate polynomials.
\end{description}  
Brown's heuristic is the cheapest, checking only simple properties of the input.  Ndrr is the most expensive (requiring real root isolation) but is the only one to consider the real geometry.

All three heuristics may identify more than one variable ordering as a suitable choice.  In this case we took the heuristic's choice to be the first of these after they had been ordered lexicographically (when written as a tuple in the reverse order to which they are projected).

\section{Experiment and results}

We ran a machine learning experiment involving these heuristics \cite{HEWDPB14}.  Two experiments were undertaken, one for CAD and another for QE by CAD, in both cases implemented by \textsc{Qepcad-B}\footnote{Interactive command-line program, freely available from \texttt{http://www.usna.edu/CS/$\sim$qepcad/B/QEPCAD.html}.}.  
7001 three-variable QE problems were taken from the nlsat database\footnote{Benchmarks for solving nonlinear arithmetic freely available from \texttt{http://cs.nyu.edu/$\sim$dejan/nonlinear/}.},
all of which were fully existential (satisfiability or SAT problems). Removing all quantifiers gave a corresponding problem set for evaluating CAD alone.  In Huang \textit{et al}. \cite{HEWDPB14} the problems were split into training, evaluation and test sets but here we report on the performance of the heuristics for all problems.  
In each case the heuristic's selections were compared according to the number of cells produced  (as opposed to computation time: so the experiment concerns the CAD theory rather than just the \textsc{Qepcad} implementation).   Note that cell counts are usually smaller for quantified problems as partial CAD techniques can be used to stop the lifting process early when the outcome is already determined.

We first observed for how many problems (and thus for what percentage of the problem set) each heuristic made the most competitive selection of the three:
\begin{center}
\begin{tabular}{lcccc}
                      & sotd			& ndrr				& Brown	  			\\
\midrule 
Quantifier free       & 4221 (60.29\%)	& 3620 (51.71\%)  	& 4523 (64.61\%)	\\
Quantified            & 4603 (65.75\%)	& 4000 (57.13\%)  	& 5166 (73.79\%)    \\
\end{tabular}
\end{center}
Hence we see that Brown's heuristic is most likely to make the best choice for our problems, both when quantified and when quantifier free.
We next investigate how much of a cell count saving is offered by each heuristic.
We made the following calculations for each problem:
\vspace*{-10pt}
\begin{shortenumerate}[(1)]
\item The average cell count of the six orderings;
\item The difference between the cell count for each heuristic's pick and the problem average;
\item The value of (2) as a percentage of (1).
\end{shortenumerate}
\vspace*{-10pt}
These calculations were made for all problems in which no variable ordering timed out (5262 of the quantifier free problems and 5332 of the quantified problems).  The data is visualised in the plot below, where the boxes indicate the second and third quartiles.  The mean and median values are given below (and marked on the plots with circles and lines respectively). 
\begin{center}
\begin{tabular}{lcccccc}
                      & \multicolumn{3}{c}{Mean average} & \multicolumn{3}{c}{Median value} \\
\cmidrule(lr){2-4}
\cmidrule(lr){5-7}
                      &	sotd 	& ndrr 		& Brown	  &	sotd 	& ndrr 	 & Brown    \\
\midrule
Quantifier free       & 27.32\% & -0.20\% 	& 25.28\% & 29.47\% & 0.00\% & 32.28\%	\\
Quantified            & 19.47\% & 4.15\%   	& 21.03\% & 14.68\% & 0.00\% & 16.67\%  \\
\end{tabular}
\end{center}
While Brown's heuristic makes the best choice the most frequently, for the quantifier free problems the average saving of using sotd is actually higher. 

\begin{center}
\includegraphics[width=0.75\textwidth]{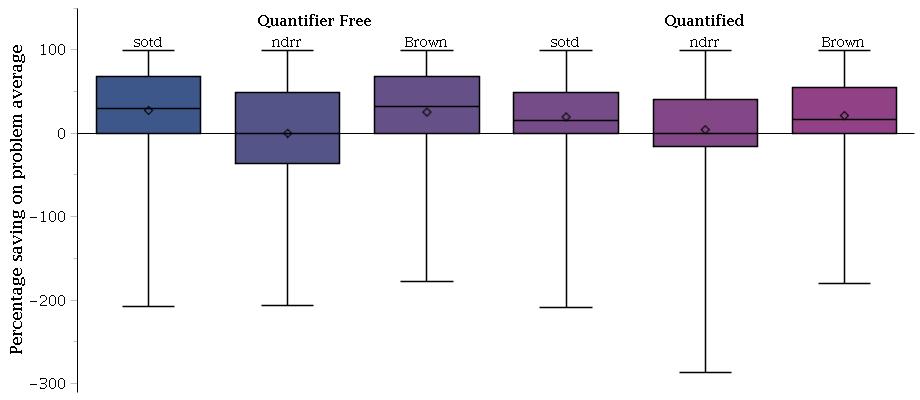}
\end{center}
 
Ndrr performs the worst on average, but there are classes of problems where it makes a better choice than the others (see Huang \textit{et al}. \cite{HEWDPB14}). For example, consider the remaining problems (those where at least one ordering timed out).  The following table describes how often each heuristic avoids a time out.  We see that for quantified problems ndrr does the best.
\begin{center}
\begin{tabular}{lccc}
                      & sotd	& ndrr  & Brown \\
\midrule 
Quantifier free       & 559		& 537   & 594	\\
Quantified            & 512     & 530   & 478   \\
\end{tabular}
\end{center}

\section{Conclusions}

We note first that the conclusion of which heuristic is the best varies depending on the criteria used to judge, whether we consider quantified problems or not (and indeed various other features of the problem which is why machine learning was useful \cite{HEWDPB14}).  
We highlight the strong performance of Brown's heuristic, surprising both because it requires the least computation and since it was not formally published.  (To the best of our knowledge it is mentioned only in notes to a tutorial at ISSAC 2004 \cite{Brown2004}).  
The problems in our example set are all from genuine applications, but are quite different to those for which CAD is normally used.  Hence further experimentation would be beneficial to see if the results can be verified more generally.  Other future work will include the testing of greedy and combination heuristics, perhaps leading to the development of a new heuristic.


\end{document}